\documentclass[letterpaper,twocolumn,10pt]{article}
\usepackage{graphicx}
\usepackage{flushend}
\usepackage{authblk}
\usepackage[english]{babel}
\usepackage{multirow, array} 

\usepackage{multicol}
\usepackage{cite}
\date{}

\begin{document}

\title{Measurement of coupled spatiotemporal coherence of parametric down-conversion under negative group velocity dispersion}

\author[1,2,*]{Paula Cutipa}
\author[1,2]{Kirill Yu. Spasibko}
\author[1,2,3]{Maria V. Chekhova}

\affil[1]{Max Planck Institute for the Science of Light, Staudtstra{\ss}e. 2, Erlangen D-91058, Germany}
\affil[2]{University of Erlangen-Nuremberg, Staudtstra{\ss}e. 7/B2, 91058 Erlangen, Germany}
\affil[3]{Physics Department, Moscow State University, Leninskiye Gory 1-2, Moscow 119991, Russia}

\affil[*]{Corresponding author: paula.cutipa@mpl.mpg.de}

\twocolumn[
\begin{@twocolumnfalse}
\maketitle
\vspace*{-1cm}
\begin{center}\rule{0.9\textwidth}{0.1mm} \end{center}
\normalsize We present a direct measurement of the spatiotemporal coherence of parametric down-conversion in the range of negative group-velocity dispersion. In this case, the frequency-angular spectra are ring-shaped and temporal coherence is coupled to spatial coherence. Correspondingly, the lack of coherence due to spatial displacement can be compensated with the introduction of time delay. We show a simple technique, based on a modified Mach-Zehnder interferometer, which allowed us to measure time coherence and near-field space coherence simultaneously, with complete control of both variables.  This technique will be also suitable for the measurement of second-order coherence, where the main applications are related to the two-photon spectroscopy.
\begin{center}\rule{0.9\textwidth}{0.1mm} \end{center}
\vspace*{0.5cm}
\end{@twocolumnfalse}
]


\maketitle
Coherence is an important characteristic of light sources, including the ones producing non-classical light ~\cite{loudon1980non,goodman2015statistical,allevi2014coherence}.  First-order coherence of a source of light is usually measured in an interferometer, by splitting the field in two (or more) parts and then overlapping the fields again, after a spatial displacement or temporal delay introduced between them. 
 
The quantitative description of coherence is given by the first-order correlation function (CF)~\cite{mandel1995optical,strekalov2005relationship}, 
\begin{equation}
G^{(1)}(t_{1},t_{2},r_{1},r_{2})=\langle E^{*}(t_1,r_1)E(t_2,r_2) \rangle,
\end{equation}
where $E(t_i,r_i)$ is the electric field at point $r_i$ and time $t_i$, $i=1,2$.
 
When the field is spatially uniform and stationary, the first-order CF depends only on the time delay and space displacement but not on the absolute values of time and space: $G^{(1)}(t_{1},t_{2},r_{1},r_{2})=G^{(1)}(\tau,\xi) $, where  $\tau=t_1-t_2$ and $\xi=r_1-r_2$ are the time delay and space shift between the two fields. The standard way to measure the first-order CF in space and time is by using, respectively, the Young and Michelson interferometers ~\cite{goodman2015statistical}. 
 
Temporal and spatial first-order correlation functions are related, respectively, to the frequency spectrum $S(\omega)$ and the transverse wavevector spectrum $S(k)$ through Fourier transforms.  These relations are known as the Wiener-Khinchine and van Cittert-Zernike theorems ~\cite{goodman2015statistical}. Usually, the frequency -- transverse wavevector spectrum is factorable, $S(\omega,k)= S_{1}(\omega)S_{2}(k) $, which means that there is no coupling between temporal and spatial coherence and they can be studied separately. However, uncoupled spatiotemporal coherence is not always the case. Indeed, it has been demonstrated that the radiation produced through parametric down-conversion (PDC) has a non-factorable frequency-wavevector spectrum $S(\omega,k) \neq  S_{1}(\omega)S_{2}(k) $~\cite{jedrkiewicz2007x,brambilla2012disclosing}.
This leads to coupled spatiotemporal coherence described by the generalization of the Wiener-Khinchin theorem, relating the first-order space-time correlation function with the frequency -- transverse wavevector spectrum \cite{picozzi2002skewed,jedrkiewicz2006emergence}: 

\begin{equation}
G^{(1)}(\tau,\xi)= \int \int  S(\omega,k) e^{ik\xi-i\omega\tau} d\tau d\xi. 
\label{eq:g1-double-integral}
\end{equation}
This property was reported experimentally for PDC in the range of positive group-velocity dispersion (GVD)~\cite{jedrkiewicz2007x}, where the coupling between temporal and spatial coherence appears due to the `X-shaped' frequency -- transverse wavevector spectrum.
The interesting consequence of the coupling between spatial and temporal coherence is that the reduction of coherence due to the spatial displacement can be compensated by the temporal delay and vice versa, which has been indeed demonstrated in Ref.~\cite{jedrkiewicz2006emergence}. However, the exact shape of $G^{(1)}(\tau,\xi)$ was not measured. Meanwhile, in the range of negative GVD ($\frac{d^2 k}{d\omega^2}<0$), the frequency-wavevector spectrum is very different. It becomes ring-shaped in the noncollinear nondegenerate case and spot-shaped in the  collinear degenerate case~\cite{strekalov2005quantum,spasibko2016ring}. Accordingly, coupled spatiotemporal coherence should be observed in this case as well, but the CF should be different. In particular, for a ring-shaped spectrum there is no collinear or degenerate emission. This feature should lead to a narrower first-order CF than in the X-shape case. In turn, the second-order CF should be also narrower in this case, which suggests very high spatial and temporal resolution in two-photon spectroscopy~\cite{dorfman2016nonlinear,spasibko2016ring,jedrkiewicz2012experimental,boitier2013two}.
 
In this work we directly measure the spatiotemporal first-order CF $G^{(1)}(\tau,\xi)$ for PDC in the negative GVD range. To this end, we develop a practical technique for measuring time and near-field space variables simultaneously. The measurement reveals coupled spatiotemporal coherence with typical coherence time on the order of femtoseconds and coherence radius on the order of micrometers. The same technique will be valid for the measurement of the second-order CF, with applications related to two-photon absorption spectroscopy \cite{leontemperature,maclean2018direct,boitier2013two}.

\begin{figure}[htbt]
	\centering
	\includegraphics[width=1\linewidth]{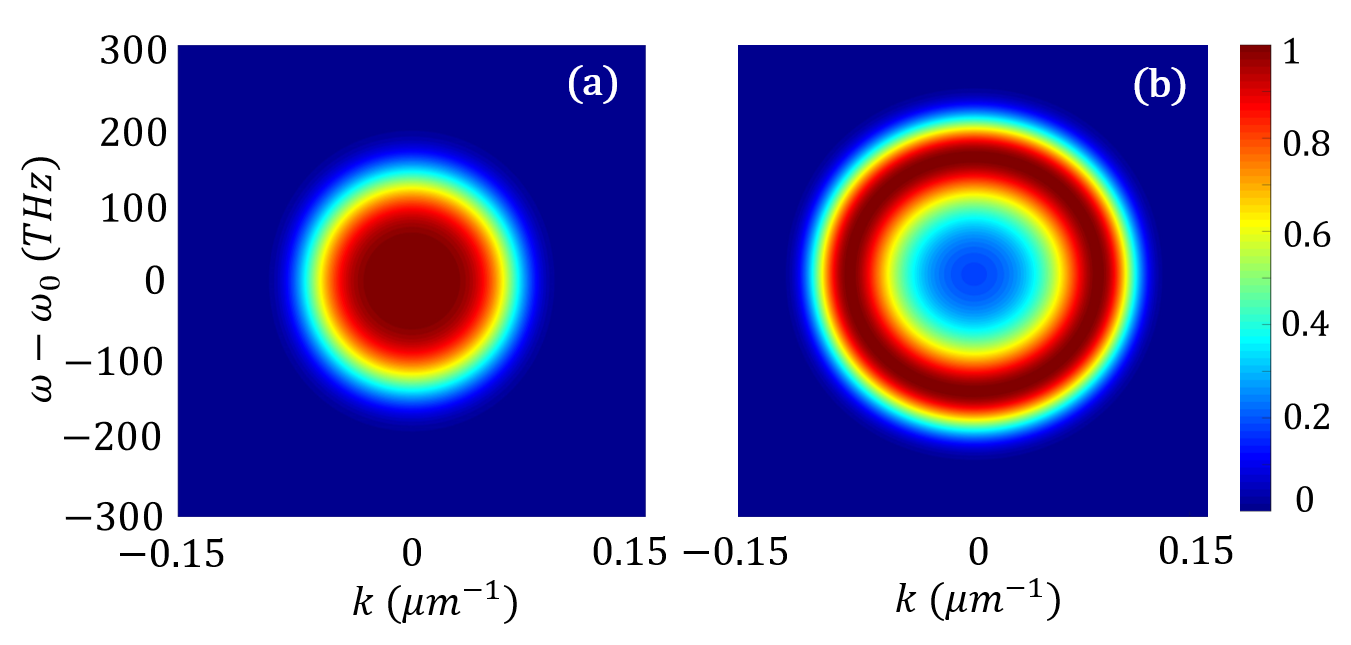}
	\caption{Calculated spectra $S(\omega,k)$ for PDC in a type-I BBO crystal pumped at $800$ nm in (a) collinear degenerate (spot-shaped) and (b) noncollinear nondegenerate (ring-shaped) cases.}
	\label{fig:theory-S(w,k)}
\end{figure}

The frequency-wavevector spectrum of both high-gain and low-gain PDC can be calculated from the longitudinal phase mismatch $ \Delta k(\omega,k)$ using the equation~\cite{brambilla2004simultaneous,brambilla2012disclosing,spasibko2012spectral}
\begin{equation}
S(\omega,k)= \left( \frac{G}{\mathcal{G}(\omega,k)} \sinh\mathcal{G}(\omega,k)  \right)  ^2,
\label{eq:sinh}
\end{equation}
where
\begin{equation}
\mathcal{G}(\omega,k) = \sqrt{G^2-\frac{(\Delta k(\omega,k)L)^2}{4} }
\label{eq:gain}
\end{equation}
is the parametric gain and $G$ is proportional to the effective second-order susceptibility and the pump field amplitude. Equation \ref{eq:sinh} was used to calculate the PDC spectra in the range of negative GVD (Fig.~\ref{fig:theory-S(w,k)}). For the calculation we assumed that PDC is pumped at $800$ nm and the crystal is a $10$ mm $\beta$-barium-borate (BBO) cut for type-I phase matching. In the collinear frequency-degenerate case, the spectrum $S(\omega,k)$ looks like a round spot as shown in Fig.~\ref{fig:theory-S(w,k)} (a).  It is broadband because the degenerate wavelength $1600$ nm is close to zero-dispersion wavelength~\cite{strekalov2005quantum}. 

When the configuration changes to the noncollinear-nondegenerate one, the spectrum becomes nonfactorable and has a typical ring-like shape (Fig.~\ref{fig:theory-S(w,k)} (b)). The size of the ring depends on the crystal orientation, i.e., the angle between its optic axis and the pump, and can be made very large. According to Eq.~\ref{eq:g1-double-integral}, in this case one can expect a non-factorable first-order CF with coherence time on the order of a few femtoseconds and coherence radius on the order of a few micrometers.

\begin{figure}[htbt] 
\centering
\includegraphics[width=1\linewidth]{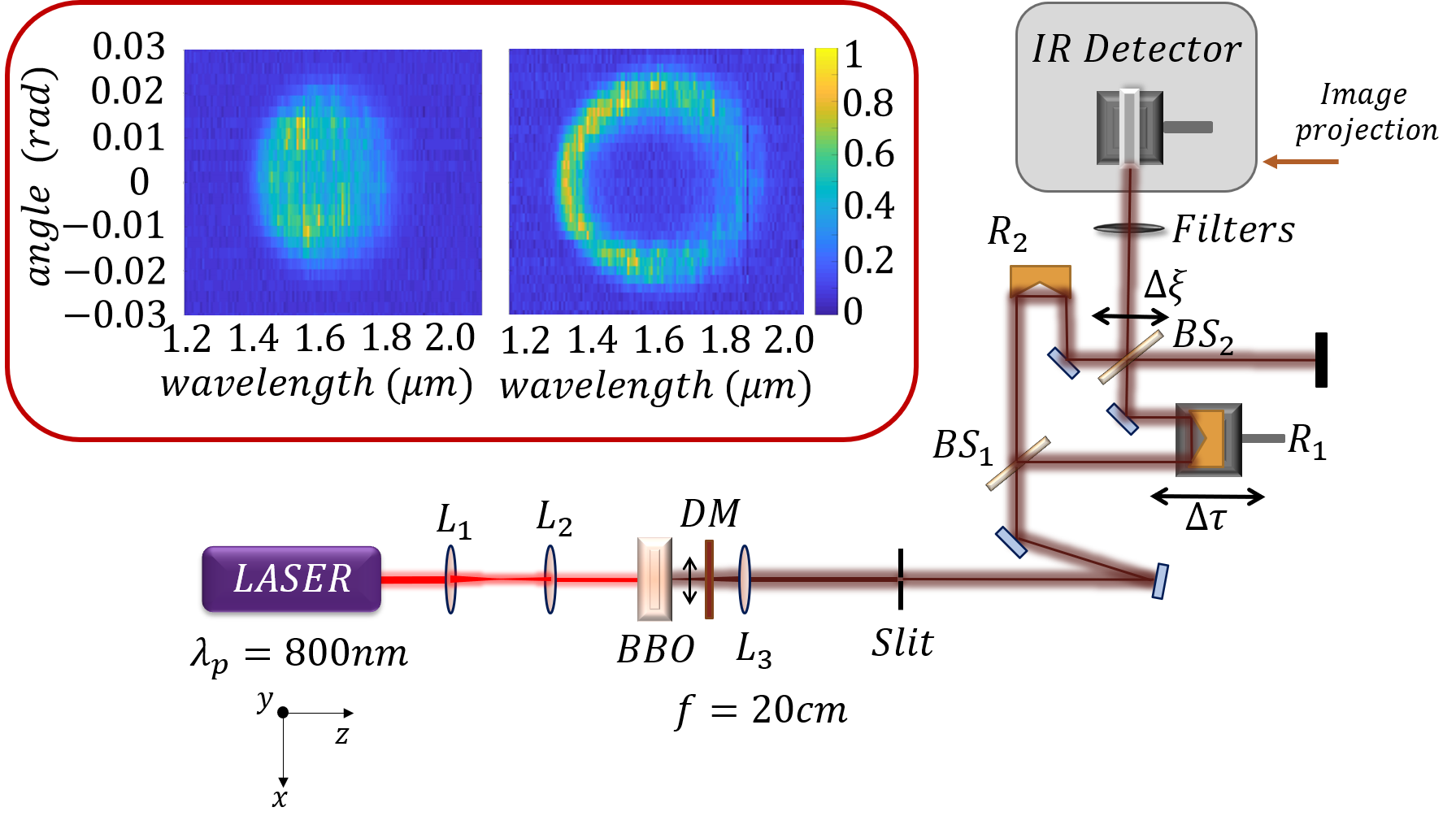}
\caption{Experimental setup for measuring the first-order correlation function. Time delay is introduced with the retroreflector $R_1$ and the space shift, with the beam splitter $BS_2$.}
\label{fig:Experimental-setup}
\end{figure}

The experimental setup for the CF measurement is shown in Fig.~\ref{fig:Experimental-setup}. It is based on a modified  Mach-Zehnder interferometer, enabling both a time delay and a space displacement between the two arms.  To pump PDC, we used an amplified Ti-sapphire laser at $800$ nm with a $1$ W mean power, $1.6$ ps pulse duration and $5$ kHz repetition rate.  The beam size was reduced by a two-lens telescope, $L_1$ and $L_2$, to a waist of $600 \mu$m. Type-I  PDC was generated in an $L= 10$ mm BBO crystal. The parametric gain at zero phase mismatch was $G=6$. For the pump incident at $19.87^{o}$ to the optic axis, exact phase matching was satisfied for collinear frequency-degenerate PDC. The optic axis was in the y-z plane; the pump beam was polarized along the y-direction (extraordinarily), as shown by a dot, while the down-converted radiation, along the x-direction (ordinarily), as shown by an arrow. For avoiding the spatial walk-off effects in the spectrum \cite{perez2015giant}, all measurements were made with a $1.6$ mm slit along the x-direction placed in the Fourier plane of the lens $L_3$ with focal length $20$ cm. This reduced the angular spectra to one-dimensional, in the plane free from the walk-off. The pump was rejected using a dichroic mirror DM, and the down-converted beam was sent to the interferometer.

The inset of Fig.~\ref{fig:Experimental-setup} shows the experimental wavelength-angular spectra $S(\lambda,\theta_{ext})$ for the collinear degenerate (a) and noncollinear nondegenerate (b) cases. The angles $\theta_{ext}$ are external, i.e., measured outside of the crystal, and are related to the transverse wavevector $k$ as $\theta_{ext}=k\lambda/(2\pi)$. These spectra correspond to the theoretical frequency-wavevector distributions $S(\omega,k)$ shown in Fig.~\ref{fig:theory-S(w,k)}. The spectra were measured the same way as in Ref.~\cite{spasibko2016ring}: by scanning in the far field the tip of a multimode fiber connected to a spectrometer, and recording the spectrum for each position of the fiber. The spectra correspond to the crystal orientation of $19.87^{o}$ (left) and $19.94^{o}$ (right). 

The PDC beam was fed into the interferometer, consisting of two thin $50:50$ beam splitters $BS_1$, $BS_2$ and two  retroreflectors $R_1$, $R_2$. The time delay was varied by moving retroreflector $R_1$ and the space displacement, by shifting the second beam splitter $BS_2$. After the interferometer, the PDC beam passed through additional filters to remove the residual pump radiation, and was registered by a home-made infrared detector. The detector was based on an InGaAs PIN photodiode with a photosensitive area of $0.3$ mm and a cutoff wavelength of $2.1\,\mu$m, which allowed us to measure the correlation properties without cutting the spectrum. For measuring the CF in the near field, we projected the output face of the crystal on the detector with lens $L_3$, placed at a distance of $43.5 \pm 0.1$ cm from the BBO crystal. The resulting image had a magnification of $6.6$. 

\begin{figure}[btth]
	\centering
	\includegraphics[width=0.9\linewidth]{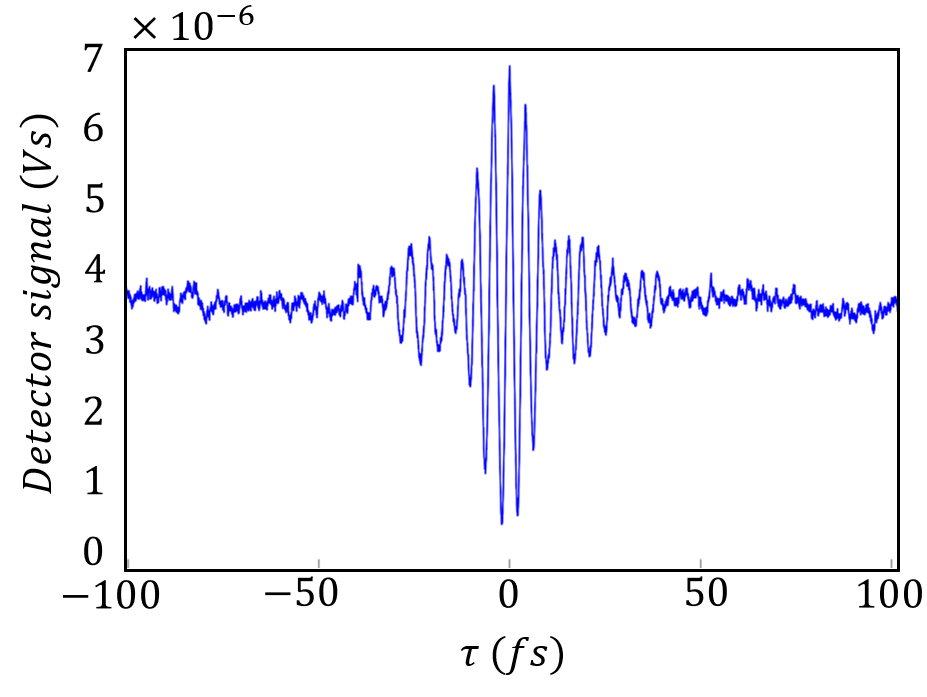}
	\caption{The detector signal measured versus the time delay in the interferometer. The crystal orientation is $19.94^o $. The fringe period corresponds to the wavelength $1600$ nm.} \label{fig:spacemagnification}
\end{figure}

The absolute value of the normalized first-order CF $g^{(1)}(\tau,\xi)\equiv G^{(1)}(\tau,\xi)/I$, with $I$ standing for the mean intensity, was measured as the interference visibility \cite{klyshko2011physical}. The latter was found by measuring the intensity at the output of the interferometer with the detector, whose sensitive area was small enough not to average over the interference fringes. 

First, we measured the output intensity versus the time delay in the interferometer, with no space shift ($BS_2$ fixed at the position of perfect alignment). The result is plotted  in Fig.~\ref{fig:spacemagnification}. The trace shows pronounced interference fringes, their period corresponding exactly to the degenerate wavelength $1600$ nm. The absolute value of normalized first-order CF $|g^{(1)}(\tau)|$  was found as the visibility of this interference pattern.

\begin{figure}[htbt]
	\centering
	\includegraphics[width=0.9\linewidth]{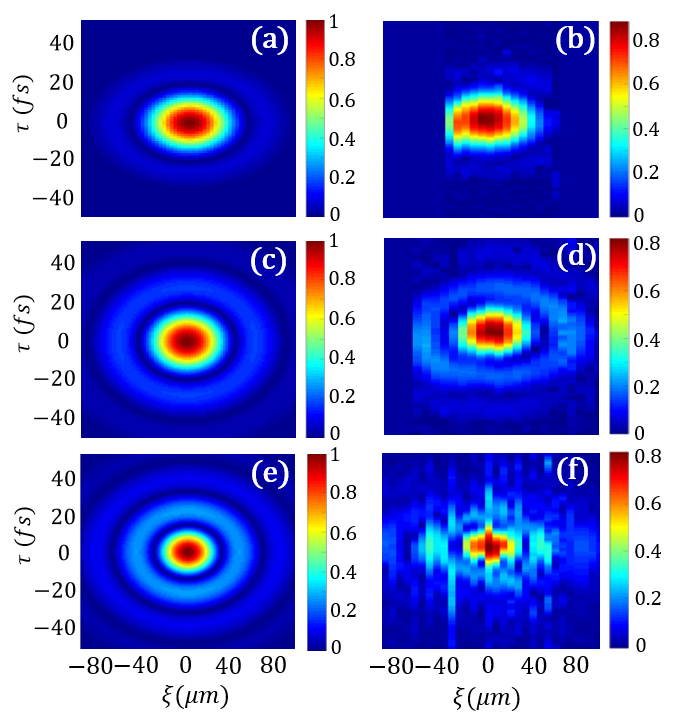}
	\caption{Calculated (a,c,e) and measured (b,d,f) distributions of $| g^{(1)}(\tau,\xi)|$ for the collinear degenerate (a,b) and noncollinear nondegenerate (c,d,e,f) PDC. The crystal orientations are $19.87^o $ (a,b), $19.90^o$ (c,d) and $19.94^o$ (e,f).} \label{fig:spatial-time-coherence}
\end{figure}

To perform the measurement of $|g^{(1)}(\tau,\xi)|$, the temporal delay and spatial displacement in the interferometer should be varied independently. Although the displacement of beamsplitter $BS_2$ introduces at the same time a temporal delay, this could be taken into account. To this end, we moved $BS_2$ in steps of $40\mu$m, and for each position the retroreflector $R_{1}$ was scanned with an automated translation stage. Each time, a profile as shown in Fig.~\ref{fig:spacemagnification} was measured, and shifted with respect to $\tau$ depending on the $BS_2$ position. Consequently applying this method, we were able to build the 2-D plot of $|g^{(1)}(\tau,\xi)|$ in time and space, as shown in Fig.~\ref{fig:spatial-time-coherence}. 
The measurements of $| g^{(1)}(\tau,\xi)|$ were performed for three crystal orientations, corresponding to the collinear degenerate and two noncollinear nondegenerate cases of PDC. 

In Fig.~\ref{fig:spatial-time-coherence}, the left panels show the calculated plots of $| g^{(1)}(\tau,\xi)|$, whereas  the right panels present the corresponding measured distributions. For the collinear degenerate case we use the crystal orientation $19.87^o $ (a,b) and for the noncollinear nondegenerate case, $19.90^o $ (c,d) and $19.94^o$ (e,f). 
The peak at the center of all distributions indicates the trivial case of $\tau=0$ and $\xi=0$, where high coherence is evidenced. Experimentally, the visibility achieved is higher than the $80\%$ for all the cases, which is good enough for the observation of the coupled  coherence exhibited in the weaker rings surrounding the central peak. For these rings, the lack of spatial coherence can be obviously compensated by the temporal delay and vice versa. 

\begin{figure}[btth]
	\centering
	\includegraphics[width=1\linewidth]{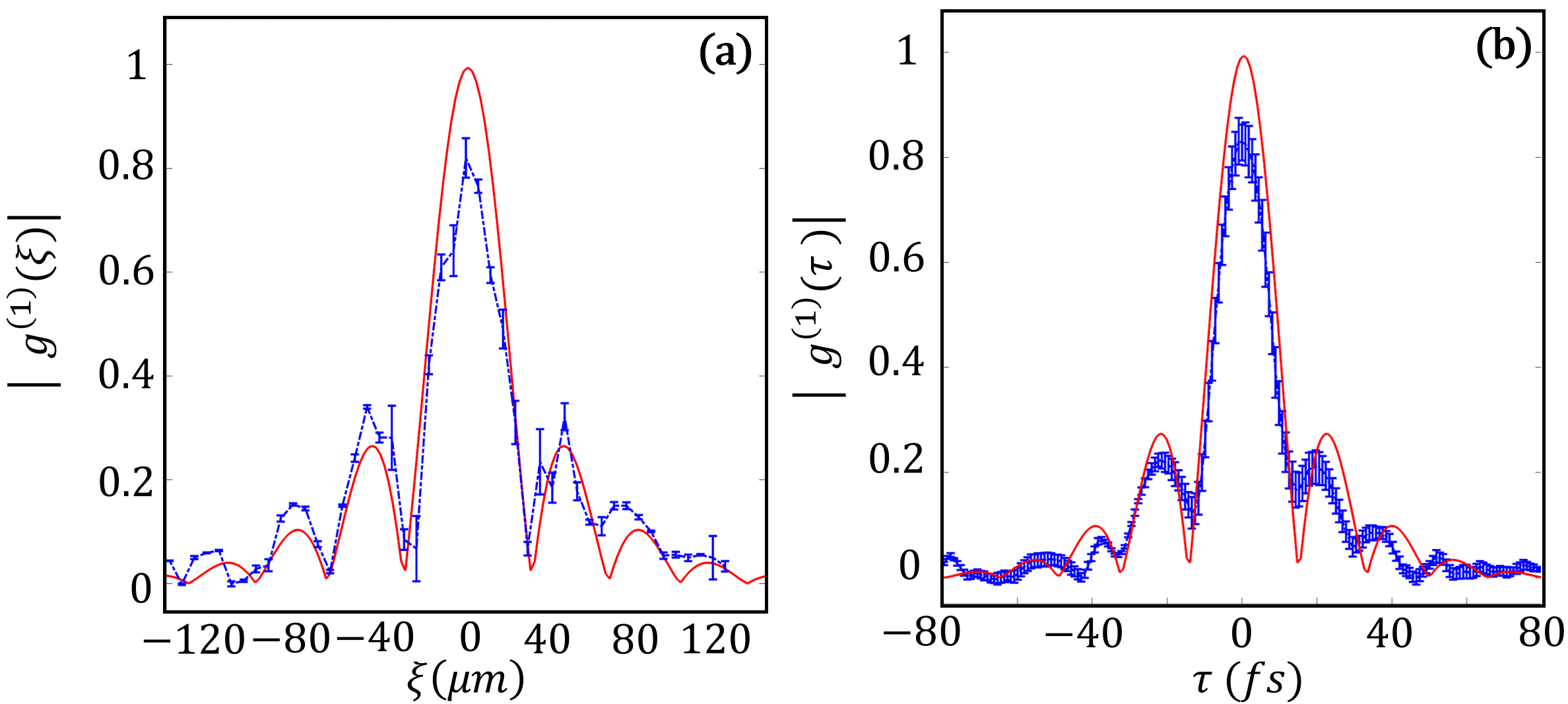}
	\caption{Experimental (blue) and calculated (red) one-dimensional `cuts' of $| g^{(1)}(\tau,\xi)|$ from Fig.~\ref{fig:spatial-time-coherence} (f): (a) $| g^{(1)}(\xi)|$ at $\tau=0$ and (b) $| g^{(1)}(\tau)|$ at $\xi=0$.} \label{fig:cuts}
\end{figure}

Figure~\ref{fig:cuts} shows one-dimensional profiles of $| g^{(1)}(\tau,\xi)|$ from Fig.~\ref{fig:spatial-time-coherence} (f): (a) as a function of $\xi$ with fixed $\tau=0$ and (b) as a function of $\tau$ with fixed $\xi=0$.
The full widths at half maximum (FWHM) in these plots correspond to the coherence radius $\xi_c$ and coherence time $\tau_c$, respectively. The obtained values are $\xi_c=38\pm7\,\mu$m and  $\tau_c=16 \pm 2$ fs, in good agreement with the theoretical numbers (see Table~\ref{Table_coh}, where the experimental and theoretical coherence parameters are listed for all three studied orientations of the crystal).

From Table~\ref{Table_coh} we see that the coherence time and radius strongly depend on the crystal orientation. In particular, in the most noncollinear nondegenerate regime (large ring-shaped spectrum) the shortest coherence time and radius are achieved. This behaviour follows from the Fourier relation between $S(k,\omega)$ and $G^{(1)}(\xi,\tau)$: the larger the ring-shaped spectrum, the shorter the coherence time and radius. In the case of the shortest coherence time and radius, as in panel (f) of Fig.~\ref{fig:spatial-time-coherence}, the finite resolution of the measurement of $1$ fs and $ 6  \mu $m reduced the quality of the distribution. 

\begin{table}[h]
\renewcommand{\arraystretch}{1.25}
 \setlength{\tabcolsep}{0.4\tabcolsep}
\begin{tabular}{|c||c|c||c|c|}
\hline
 $\theta$ & $\tau_c$, exp & $\tau_c$, theory & $\xi_c$, exp & $\xi_c$, theory\\
\hline
 $19.87^o$   & $36 \pm 2$ fs & $28$ fs  & $67\pm 8 \, \mu$m & $59\, \mu$m\\
\hline
 $19.90^o$   & $19 \pm 2$ fs & $22$ fs & $48\pm 12 \mu$m & $46\,\mu$m\\
\hline
 $19.94^o$   & $16 \pm 2$ fs & $17$ fs & $38\pm 7 \mu$m & $37\,\mu$m\\
\hline
\end{tabular}
\caption{Coherence times and radiuses of PDC measured and calculated for different crystal orientations ($\theta$) of the BBO crystal} \label{Table_coh}
\renewcommand{\arraystretch}{1}
\end{table}

Another feature noticeable in Fig.~\ref{fig:spatial-time-coherence} is that the rings surrounding the central peak get more and more pronounced as the orientation gets further from collinear degenerate and the ring-shaped spectra become larger. Indeed, the height of the first ring is about $0.22$ for orientation $19.90^o$ and $0.30$ for orientation $19.94^o$. This again demonstrates that the coupling between temporal and spatial coherence occurs due to the non-factorability of the frequency - transverse wavevector spectrum.

In conclusion, we have demonstrated the measurement of coupled spatiotemporal coherence, with the simultaneous control over space and time variables. We applied this technique to the radiation of high-gain parametric down-conversion in the negative GVD range, where the frequency-wavevector spectra are ring-shaped. The obtained 2D distributions of the first-order correlation function demonstrate strong coupling between temporal and spatial degrees of freedom. Due to the large frequency and angular bandwidth of the radiation, and especially due to the absence of collinear frequency-degenerate emission, it is possible to achieve very short time and space coherence scales (on the order of femtoseconds and micrometers, respectively).  The same experimental technique will be suitable for measuring the second-order correlation function and therefore implementing two-photon space-time nonlinear spectroscopy with PDC radiation. 

We thank Denis Kopylov for the help in experiment.

\section*{Funding Information}
P. Cutipa acknowledges the funding provided by the CONICYT PFCHA/DOCTORADO BECAS CHILE/2019 - 72180453.



\bibliography{referencias}

\begin{thebibliography}{10}

\bibitem{loudon1980non}
Rodney Loudon.
\newblock Non-classical effects in the statistical properties of light.
\newblock {\em Reports on Progress in Physics}, 43(7):913, 1980.

\bibitem{goodman2015statistical}
Joseph~W Goodman.
\newblock {\em Statistical optics}.
\newblock John Wiley \& Sons, 2015.

\bibitem{allevi2014coherence}
Alessia Allevi, Ottavia Jedrkiewicz, Enrico Brambilla, Alessandra Gatti,
  J~Pe{\v{r}}ina~Jr, O~Haderka, and Maria Bondani.
\newblock Coherence properties of high-gain twin beams.
\newblock {\em Physical Review A}, 90(6):063812, 2014.

\bibitem{mandel1995optical}
Leonard Mandel and Emil Wolf.
\newblock {\em Optical coherence and quantum optics}.
\newblock Cambridge university press, 1995.

\bibitem{strekalov2005relationship}
Dmitry Strekalov, Andrey~B Matsko, Anatoliy~A Savchenkov, and Lute Maleki.
\newblock Relationship between quantum two-photon correlation and classical
  spectrum of light.
\newblock {\em Physical Review A}, 71(4):041803, 2005.

\bibitem{jedrkiewicz2007x}
Ottavia Jedrkiewicz, Matteo Clerici, Antonio Picozzi, Daniele Faccio, and Paolo
  Di~Trapani.
\newblock X-shaped space-time coherence in optical parametric generation.
\newblock {\em Physical Review A}, 76(3):033823, 2007.

\bibitem{brambilla2012disclosing}
Enrico Brambilla, Ottavia Jedrkiewicz, Luigi~Alberto Lugiato, and Alessandra
  Gatti.
\newblock Disclosing the spatiotemporal structure of parametric down-conversion
  entanglement through frequency up-conversion.
\newblock {\em Physical Review A}, 85(6):063834, 2012.

\bibitem{picozzi2002skewed}
Antonio Picozzi and Marc Haelterman.
\newblock Skewed coherence ong space-time trajectories in parametric generation
  processes.
\newblock In {\em Nonlinear Guided Waves and Their Applications}, page NLMB7.
  Optical Society of America, 2002.

\bibitem{jedrkiewicz2006emergence}
Ottavia Jedrkiewicz, Antonio Picozzi, Matteo Clerici, Daniele Faccio, and Paolo
  Di~Trapani.
\newblock Emergence of x-shaped spatiotemporal coherence in optical waves.
\newblock {\em Physical review letters}, 97(24):243903, 2006.

\bibitem{strekalov2005quantum}
Dmitry Strekalov, Andrey~B Matsko, Anatoly Savchenkov, and Lute Maleki.
\newblock Quantum-correlation metrology with biphotons: where is the limit?
\newblock {\em Journal of Modern Optics}, 52(16):2233--2243, 2005.

\bibitem{spasibko2016ring}
Kirill~Yu Spasibko, Denis~A Kopylov, Tatiana~V Murzina, Gerd Leuchs, and
  Maria~V Chekhova.
\newblock Ring-shaped spectra of parametric downconversion and entangled
  photons that never meet.
\newblock {\em Optics letters}, 41(12):2827--2830, 2016.

\bibitem{dorfman2016nonlinear}
Konstantin~E Dorfman, Frank Schlawin, and Shaul Mukamel.
\newblock Nonlinear optical signals and spectroscopy with quantum light.
\newblock {\em Reviews of Modern Physics}, 88(4):045008, 2016.

\bibitem{jedrkiewicz2012experimental}
O~Jedrkiewicz, A~Gatti, E~Brambilla, and P~Di~Trapani.
\newblock Experimental observation of a skewed x-type spatiotemporal
  correlation of ultrabroadband twin beams.
\newblock {\em Physical review letters}, 109(24):243901, 2012.

\bibitem{boitier2013two}
Fabien Boitier, Antoine Godard, Nicolas Dubreuil, Philippe Delaye, Claude
  Fabre, and Emmanuel Rosencher.
\newblock Two-photon-counting interferometry.
\newblock {\em Physical Review A}, 87(1):013844, 2013.

\bibitem{leontemperature}
Roberto de~J Le{\'o}n-Montiel, Ji{\v{r}}{\'\i} Svozil{\'\i}k, Juan~P Torres,
  and Alfred~B U’Ren.
\newblock Temperature-controlled entangled-photon absorption spectroscopy.
\newblock {\em Physical review letters}, 123(2):023601, 2019.

\bibitem{maclean2018direct}
Jean-Philippe~W MacLean, John~M Donohue, and Kevin~J Resch.
\newblock Direct characterization of ultrafast energy-time entangled photon
  pairs.
\newblock {\em Physical review letters}, 120(5):053601, 2018.

\bibitem{brambilla2004simultaneous}
Enrico Brambilla, Alessandra Gatti, Morten Bache, and Luigi~A Lugiato.
\newblock Simultaneous near-field and far-field spatial quantum correlations in
  the high-gain regime of parametric down-conversion.
\newblock {\em Physical Review A}, 69(2):023802, 2004.

\bibitem{spasibko2012spectral}
K~Yu Spasibko, T~Sh Iskhakov, and Maria~V Chekhova.
\newblock Spectral properties of high-gain parametric down-conversion.
\newblock {\em Optics express}, 20(7):7507--7515, 2012.

\bibitem{perez2015giant}
Angela~M P{\'e}rez, Kirill~Yu Spasibko, Polina~R Sharapova, Olga~V Tikhonova,
  Gerd Leuchs, and Maria~V Chekhova.
\newblock Giant narrowband twin-beam generation along the pump-energy
  propagation direction.
\newblock {\em Nature communications}, 6(1):1--5, 2015.

\bibitem{klyshko2011physical}
David~Nikolaevich Klyshko, Maria Chekhova, and Sergey Kulik.
\newblock {\em Physical foundations of quantum electronics}.
\newblock World Scientific, 2011.

\end{thebibliography}

\end{document}